\documentclass[11pt]{article}
\usepackage{amsmath}
\usepackage{amssymb}
\usepackage{epsfig}
\usepackage{wrapfig}
\usepackage{times}
\begin{document}

\newcommand{\sss}{\scriptscriptstyle}
\def\shalf{\hbox{${\textstyle{\frac{1}{2}}}$}}
\def\quarter{\hbox{${\textstyle{\frac{1}{4}}}$}}
\def\eights{\hbox{${\textstyle{\frac{1}{8}}}$}}
\def\reals{\mathbb{R}}
\def\integers{\mathbb{Z}}
\def\overc{\hbox{${\textstyle{\frac{1}{c}}}$}}
\def\fourpioverc{\hbox{${\textstyle{\frac{4\pi}{c}}}$}}
\def\i{\rm\sss initial}
\def\f{\rm\sss final}
\setlength{\columnsep}{8pt}
\hyphenation{gravi-tational correc-tions bodies mani-fold vanish-ing analo-gous easi-ly loca-lized}

\title{Close encounters of black holes}
\author{
\small Domenico Giulini         \\ 
\small  Department of Physics    \\
\small  University of Freiburg   \\
\small  Hermann-Herder-Strasse 3 \\
\small  D-79104 Freiburg         \\
\small  Germany}
\date{}
\maketitle

\begin{abstract}\noindent
This is an introduction into the problem of how to set up black hole 
initial-data for the matter-free field equations of General Relativity.
The approach is semi-pedagogical and addresses a more general audience 
of astrophysicists and students with no specialized training in General 
Relativity beyond that of an introductory lecture.\footnote{  
This is the written version of a lecture delivered at the school 
``The Galactic Black Hole'', held between August 26.-31. in 2001 at the 
Physics Center of the German Physical Society in Bad Honnef (Germany). 
It is published in: Heino Falcke and Friedrich W Hehl (editors) 
\emph{The Galactic Black Hole (Lectures on General Relativity and 
Astrophysics)}, IOP Publishing (Bristol) 2003.} 

\end{abstract}

\section{Introduction and motivation}
\label{sec:introduction}
In my lecture I will try to explain how scattering and merging 
processes between black holes can be described analytically in 
general relativity(GR). This is a vast subject and I will focus 
attention to the basic issues, rather than trying to explain analytical 
details of approximation schemes etc. I will also not discuss numerical 
aspects, which are beyond my competence, and which would anyway 
require a separate lecture. I will address the following main topics:
\begin{enumerate}
\item[1)]
a first step beyond Newtonian gravity,
\item[2)]\vspace{-2mm}
constrained evolutionary structure of Einstein's equations,
\item[3)]\vspace{-2mm}
the 3+1- split and the Cauchy initial-value problem,
\item[4)]\vspace{-2mm}
black hole data,
\item[5)]\vspace{-2mm}
problems and recent developments,
\end{enumerate}
with emphasis on the fourth entry. However, I will also spend some 
time to explain some of the specialties of GR, like the absence of 
a point-particle concept and the non-trivial linkage between the 
field equations and the equations of motion for matter. These points 
should definitely be appreciated before one goes on to discuss black 
holes, which are solutions to the \emph{vacuum} Einstein equations 
representing \emph{extended} objects. 

The following points seem to me the main motivations for 
studying the problem of black hole collision: 
\begin{itemize}  
\item
Coalescing  black holes are regarded as promising sources 
for the detection of gravitational waves by earth-based instruments.
\item
Close encounters of black holes provide physically relevant 
situations for the investigation of the strong-field regime 
of general relativity.
\item
The dynamics of simple black hole configurations is regarded 
as ideal testbed for numerical relativity.
\end{itemize}
My conventions are as follows: space-time is a manifold $M$ with
Lorentzian metric $g$ of signature $(-,+,+,+)$. Greek indices 
range from $0$ to $3$, latin indices from $1$ to $3$ unless stated 
otherwise. The covariant derivative is denoted $\nabla_{\mu}$,
ordinary partial derivatives by $\partial_{\mu}$ or sometimes 
simply by a lower-case $,\mu$. The relation $:=$ ($=:$) defines 
the left (right) hand side. 
The gravitational constant in GR is $\kappa=8\pi G/c^2$, where 
$G$ is Newton's constant and $c$ the velocity of light. A symbol
like $O(\epsilon^n)$ stands collectively for terms falling off at 
least as fast as $\epsilon^n$.

\section{A first step beyond Newtonian gravity}
It can hardly be overstressed how useful the concept of a \emph{mass-point} 
is in Newtonian mechanics and gravity. It allows to \emph{pointwise} 
probe the gravitational field and to reduce the dynamical problem to 
the mathematical problem of finding solutions to a system of 
\emph{finitely} many \emph{ordinary} differential equations. To be sure, just 
postulating the existence of mass points is not sufficient. 
To be consistent with known laws of physics one must eventually understand 
the point mass as an idealization of a highly localized mass distribution 
which obeys known field-theoretic laws, such that in the situations at 
hand most of the field degrees of freedom effectively decouple from the 
dynamical laws for those collective degrees of freedom one is interested 
in, like e.g. the centre-of-mass.  In Newtonian gravity this usually 
requires clever approximation schemes but is not considered to be a 
problem of fundamental nature. Although this is true for the specific 
linear theory of Newtonian gravity, this needs not be so for comparably 
simple generalizations, as will become clear below.

In GR the situation is markedly different. 
A concentration of more than one Schwarzschild mass in a region of radius 
less than the Schwarzschild radius will lead to a black hole whose 
behaviour away from the stationary state usually cannot be well described 
by finitely many degrees of freedom. It shakes and vibrates, thereby 
radiating off energy and angular momentum in form of gravitational 
radiation. Moreover, it is an extended object and cannot be unambiguously 
ascribed an (absolute or relative) position or individual mass. Hence the 
problem of motion, and therefore the problem of scattering of black holes,
cannot be expected to merely consist of \emph{corrections} to Newtonian 
scattering problems. Rather, the whole kinematic and dynamical setup 
will be different where many of the established concepts of Newtonian 
physics need to be replaced or at least adapted, very often in a somewhat  
ambiguous way. Among those are mass, distance, and kinetic energy.
For example, one may try to solve the following straightforward sounding 
problem in GR, whose solution one might think has been given long ago. 
Given are two unspinning black holes, momentarily at rest,  with equal 
individual mass $m$, mutual distance $\ell$, and no initial gravitational 
radiation around. What is the amount of energy released via gravitational 
radiation during the dynamical infall? In such situation we can usually 
make sense of the notions of `spin' (hence unspinning) and `mass'; but 
ambiguities generally exist in defining `distance' and, most important
of all, `initial gravitational radiation'. Such difficulties persist 
over and above the ubiquitous analytical and/or numerical problems 
which are currently under attack by many research groups.

To those who are not so familiar with GR and like to see Newtonian 
analogies, I wish to mention that there is a way to \emph{consistently} 
model some of the non-linear features of Einstein's equations in a 
Newtonian context, which shares the property that it does not allow 
for point masses. I will briefly describe this model since it does 
not seem to be widely known. 

First recall the field equation in Newtonian gravity, which allows to 
determine the gravitational potential $\phi$ (whose negative gradient, 
-$\vec\nabla\phi$, is the gravitational field) from the mass density 
$\rho$ (the `source' of the gravitational field):
\begin{equation}
\Delta\phi=4\pi G\rho\,.
\label{eq:poisson-eq}
\end{equation}
Now suppose one imposes the following principle for a modification
of (\ref{eq:poisson-eq}): all energies, including the self-energy 
of the gravitational field, act as source for the gravitational 
field. In order to convert an energy density $\varepsilon$ into 
a mass density $\rho$, we adopt the relation $\varepsilon=\rho c^2$ 
from special relativity (the equation we will arrive at can easily 
be made Lorentz invariant by adding appropriate time-derivatives). 
The question then is whether one can modify the source term of 
(\ref{eq:poisson-eq}) such that 
$\rho\rightarrow\rho+\rho_{\sss\rm grav}$ with  
$\rho_{\sss\rm grav}:=\varepsilon_{\sss\rm grav}/c^2$, where
$\varepsilon_{\sss\rm grav}$ is the energy density of the 
gravitational field \emph{as predicted by this very same equation}
(condition of self-consistency). It turns out that there is 
indeed a unique such modification, which reads 
\begin{equation}
\Delta\phi=
\frac{4\pi G}{c^2}\,\phi\left[\rho
+\frac{c^2}{8\pi G}(\vec\nabla\phi/\phi)^2\right]\,.
\label{eq:mod-poisson-eq}
\end{equation}
It is shown in the Appendix that this equation indeed 
satisfies the `energy principle' as just stated. (For more 
information and a proof of uniqueness, see \cite{Giulini 1996}.)
The gravitational potential is now required to be always positive, 
tending to the value $c^2$ at spatial infinity (rather than zero as 
for (\ref{eq:poisson-eq})). The second term on the right hand side 
of (\ref{eq:mod-poisson-eq}) corresponds to the energy density of 
the gravitational field. Unlike the energy density following from 
(\ref{eq:poisson-eq}) (which is $-\frac{1}{8\pi G}\vert\nabla\phi\vert^2$) 
it is now \emph{positive} definite. This does not contradict 
attractivity of gravity for the following reason: the rest-energy 
density of a piece of matter is in this theory not given by $\rho c^2$,
but by $\rho\phi$, that is, it depends on the value of the 
gravitational potential at the location of matter. The same piece of 
matter located at lower gravitational potential has less energy than 
at higher potential values. In GR this is called the 
universal redshift effect. Here, as in GR, the active gravitational-mass 
also suffers from this redshift, as is immediate from the first term on 
the right hand side of (\ref{eq:mod-poisson-eq}), where $\rho$ does not 
enter alone, as in (\ref{eq:poisson-eq}), but is multiplied with the 
gravitational potential $\phi$. With respect to these features our 
modification (\ref{eq:mod-poisson-eq}) of Newtonian gravity mimics 
GR quite well.

We mention in passing that (\ref{eq:mod-poisson-eq}) can be `linearized'
by introducing the dimensionless field $\psi:=\sqrt{\phi}/c$, in terms 
of which (\ref{eq:mod-poisson-eq}) reads
\begin{equation}
\Delta\psi=\frac{2\pi G}{c^2}\rho\psi\,.
\label{eq:linearization}
\end{equation}
The boundary conditions are now $\psi(r\rightarrow\infty)\rightarrow 1$.
Hence only those linear combinations of solutions are again solutions 
whose coefficients add up to one. For $\rho\geq 0$ it also follows 
that solutions to (\ref{eq:linearization}) can never assume negative 
values, since otherwise the function $\psi$ must have a negative minimum 
(because of the positive boundary values) and therefore non-negative 
second derivatives there. But then (\ref{eq:linearization}) 
cannot be satisfied at the minimum, hence $\psi$ must be nonnegative
everywhere. This implies that solutions of (\ref{eq:mod-poisson-eq})
are also non-negative. To be sure, for mathematical purposes 
(\ref{eq:linearization}) is easier to use than (\ref{eq:mod-poisson-eq}), 
but note that $\phi$ and not $\psi$ is the physical gravitational 
potential.

We now show how these non-linear features render impossible the 
notion of point mass, and even induce a certain black-hole behaviour
on their solutions. Let us be interested in static, spherically 
symmetric solutions to (\ref{eq:mod-poisson-eq}) with source $\rho$,
which is zero for $r>R$ and constant for $r<R$. We need to distinguish 
two notions of mass. One mass just counts the amount of `stuff' located
within $r<R$. You may call it `bare mass' or `baryonic mass', since 
for ordinary matter it is proportional to the baryon number.
We denote it by $M_B$. It is simply given by 
\begin{equation} 
M_B:=\int_{\rm\sss space}d^3x\ \rho\,.
\label{eq:bare-mass}
\end{equation}
The other mass is the `gravitational mass', which is measured 
by the amount of flux of the gravitational field to `infinity', 
that is, through the surface of a sphere whose radius tends to 
infinity. We call this mass $M_G$. It is given by
\begin{equation}
M_G:=\frac{1}{4\pi G}\lim_{r\rightarrow\infty}
\int_{S^2(r)}(-\vec\nabla\phi\cdot\vec n)\, d\sigma\,,
\label{eq:grav-mass}
\end{equation}
where $r=\vert\vec x\vert$, $\vec x/r=\vec n$, $S^2(r)$ is the 2-sphere 
or radius $r$ and $d\sigma$ is its surface element. $M_G$ should be 
identified with the total inertial mass of the system, in full analogy 
to the ADM-mass in GR (see eq. (\ref{eq:ADM-mass}) below). 
Hence $M_Gc^2$ is the total energy of 
the system, with gravitational binding energy also taken into account.
The masses $M_B$ and $M_G$ can dimensionally be turned into radii by writing 
$R_B:=GM_B/c^2$ and $R_G:=GM_G/c^2$, and further turned into 
dimensionless quantities via rescaling with $R$, the radius of our 
homogeneous star. We write $x:=R_B/R$ and $y:=R_G/R$. 

For each pair of values for the two parameters $M_B$ and $R$ there is 
a unique homogeneous-star-solution to (\ref{eq:mod-poisson-eq}), whose 
simple analytical form needs not interest us here (see \cite{Giulini 1996}). 
Using it we can calculate $M_G$, whose dependence on the parameters is 
best expressed in terms of the dimensionless quantities $x$ and $y$:
\begin{equation}
y=f(x)=2\,\left(1-\frac{\tanh(\sqrt{3x/2})}
                       {\sqrt{3x/2}}\right)\,.
\label{eq:MG-MB-relation}
\end{equation}
The function $f$ maps the interval $[0,\infty]$ monotonically
to $[0,2]$. This implies the following inequality 
\begin{equation}
M_G< 2Rc^2/G
\label{eq:schwarzschild-inequality}
\end{equation}
which says that the gravitational mass of the star is bounded by a 
purely geometric quantity. It corresponds to the statement in GR 
that the star's radius  must be bigger than its Schwarzschild radius,
which in isotropic coordinates is indeed given by $R_S=GM_G/2c^2$.
It can be proven \cite{Giulini 1996} that the bound 
(\ref{eq:schwarzschild-inequality}) still exists for non-homogeneous 
spherically symmetric stars, so that the somewhat unphysical homogeneity 
assumption can be lifted. The physical reason for this inequality is the 
`redshift', i.e. the fact, that the same bare mass at lower 
gravitational potential produces less gravitational mass. Hence 
adding more and more bare mass into the same volume pushes the 
potential closer and closer to zero (recall that $\phi$
is always positive) so that the added mass becomes less 
and less effective in generating gravitational fields. The inequality 
then expresses the mathematical fact that this `redshifting' is 
sufficiently effective so as to give finite upper bounds to the 
gravitational mass, even for unbounded amounts of bare mass. 

The energy balance can also be nicely exhibited. Integrating the 
matter energy density $\phi\rho$ and the energy density 
of the gravitational field, $\frac{c^4}{8\pi G}(\vec\nabla\phi/\phi)^2$,
we obtain
\begin{eqnarray}
E_{\rm\sss matter} &=& 
M_Bc^2\,\left(1-\frac{6R_B}{5R} + O(R_B^2/R^2)\right)\,,
\label{eq:energy-matter}\\
E_{\rm\sss field}  &=& 
M_Bc^2\,\left(\frac{3R_B}{5R}   + O(R_B^2/R^2)\right)\,,
\label{eq:field-field}\\
E_{\rm\sss total}  &=& 
M_Bc^2\,\left(1-\frac{3R_B}{5R} + O(R_B^2/R^2)\right) = M_Gc^2\,.
\label{eq:energy-total}
\end{eqnarray}
Note that the term $-3M_Bc^2R_B/5R$ in (\ref{eq:energy-total}) is just 
the Newtonian binding energy. At this point it is instructive to verify 
the remarks we made above about the positivity of the gravitational 
energy. Shrinking a mass distribution \emph{enhances} the field energy,
but diminishes the matter energy twice as fast, so that the
overall energy is also diminished, as it must due to the attractivity 
of gravity. But here this is achieved with all energies involved 
being positive, unlike in Newtonian gravity. Note that the 
total energy, $M_G$, cannot become negative (since $\phi$ cannot 
become negative, as already shown). Hence one also cannot extract an 
infinite amount of energy by unlimited compression, as it is possible 
in Newtonian gravity. This is the analogue in our model theory to the 
positive mass theorem in GR.

We conclude by making the point announced above, namely that the 
inequality (\ref{eq:schwarzschild-inequality}) shows that point 
objects of finite gravitational mass do not exist in the theory based 
upon (\ref{eq:mod-poisson-eq}); mass implies extension! Taken together 
with the lesson from special relativity, that extended rigid bodies
also do not exist (since the speed of elastic waves is less than 
$c$), we arrive at the conclusion that the dynamical problem of 
gravitating bodies and their interaction is fundamentally of
field-theoretic (rather than point-mechanical) nature. Its proper 
realization is GR to which we now turn.

\section{Constrained evolutionary structure of Einstein's equations}
In GR the basic field is the spacetime metric $g_{\mu\nu}$, which 
comprises the gravitational and inertial properties of spacetime.
It defines what inertial motion is, namely a geodesic 
\begin{equation}
{\ddot x}^{\lambda}+
\Gamma^{\lambda}_{\mu\nu}{\dot x}^{\mu}{\dot x}^{\nu}=0
\label{eq:geodesic}
\end{equation}
with respect to the Levi-Civita connection
\begin{equation}  
\Gamma^{\lambda}_{\mu\nu}:=
\shalf g^{\lambda\sigma}
(-g_{\mu\nu,\sigma}+g_{\sigma\mu,\nu}+g_{\nu\sigma,\mu})\,.
\label{eq:christoffel}
\end{equation}
(Since inertial motion is `force-free' by definition, you may rightly 
ask whether it is correct to call gravity a `force'.) 
The gravitational field $g_{\mu\nu}$ is linked to the matter content 
of spacetime, represented in form of the energy momentum tensor
$T_{\mu\nu}$, by Einstein's equations 
\begin{equation}
G_{\mu\nu}:=R_{\mu\nu}-\shalf g_{\mu\nu}R=\kappa T_{\mu\nu}.
\label{eq:einstein}
\end{equation}
Due to the gauge-invariance with respect to general 
differentiable point transformations (i.e. diffeomorphisms) of 
spacetime one has the \emph{identities} (as a consequence of 
Noether's 2nd theorem) 
\begin{equation}
\nabla_{\mu}G^{\mu\nu}\equiv 0\,.
\label{eq:bianchi-id}
\end{equation}
Being `identities' they hold for any $G^{\mu\nu}$, independent of any 
field equation. With respect to some coordinate system 
$x^{\mu}=(x^0,\cdots x^3)$ we can expand (\ref{eq:bianchi-id}) in
terms of ordinary derivatives. Preferring the coordinate $x^0$, this 
reads
\begin{equation}
\partial_0G^{0\nu}=-\partial_kG^{k\nu}
                   -\Gamma^{\mu}_{\mu\lambda}G^{\lambda\nu}
                   -\Gamma^{\nu}_{\mu\lambda}G^{\mu\lambda}.
\label{eq:bianchi-expanded}
\end{equation}
Since $G^{\mu\nu}$ contains no higher derivatives of $g_{\mu\nu}$
than the second, the right-hand side of this equation also contains
only 2nd $x^0$-derivatives. Hence (\ref{eq:bianchi-expanded}) 
implies that the four components $G^{0\nu}$ only involve first 
$x^0$-derivatives. Now choose $x^0$ as time coordinate. The four 
$(0,\nu)$-components of (\ref{eq:einstein}) then do not involve second 
time derivatives, unlike the space-space components $(i,j)$. Hence 
the time-time and time-space components are \emph{constraints}, that is,
equations that constrain the allowed choices of initial data, rather 
than evolving them. 

That is not an unfamiliar situation, which similarly occurs for 
Maxwell's equations in electromagnetism (EM). Let us recall this 
analogy. We consider the four-dimensional form of Maxwell's 
equations in terms of the vector potential $A_{\mu}$, whose 
antisymmetric derivative is the field-tensor 
$F_{\mu\nu}:=\partial_{\mu}A_{\nu}-\partial_{\nu}A_{\mu}$, 
comprising the electric ($E_i=F_{0i}$) and magnetic 
($B_i=-F_{jk}$, $ijk$ cyclic) field. Maxwell's equations are 
\begin{equation}
E^{\nu}:=\partial_{\mu}F^{\mu\nu}-{\textstyle \frac{4\pi}{c}} 
j^{\nu}=0\,.
\label{eq:maxwell}
\end{equation}
Due to its antisymmetry, the field tensor obviously obeys 
the identity
\begin{equation}
\partial_{\mu}\partial_{\nu}F^{\mu\nu}\equiv 0\,,
\label{eq:em-id}
\end{equation}
which here is the analogue of (\ref{eq:bianchi-id}), an 
identity involving third derivatives in the field variables.
Using (\ref{eq:em-id}) in the divergence of (\ref{eq:maxwell}) 
yields
\begin{equation}
\partial_{\nu}E^{\nu}=-{\textstyle\frac{4\pi}{c}}
\partial_{\nu}j^{\nu}\,,
\label{eq:identity-maxwell}
\end{equation}
which shows that Maxwell's equations imply charge conservation 
as integrability condition. Let us interpret the r\^ole of charge 
conservation in the initial value problem. 
Decomposing (\ref{eq:em-id}) into space and time derivatives gives 
\begin{equation}
\partial_0\partial_{\nu}F^{0\nu}=
-\partial_k\partial_{\nu}F^{k\nu}\,.
\label{eq:em-id-dec}
\end{equation}
Again the right hand side involves only second time derivatives
implying that the $0$-component of (\ref{eq:maxwell}) involves no 
second time derivatives. Hence the time component of 
(\ref{eq:maxwell}) is merely a constraint on the initial data;
clearly it is just Gauss' law $\vec\nabla\cdot\vec E-4\pi\rho=0$.
Its change under time evolution according to Maxwell's 
equations is  
\begin{eqnarray}
\partial_0E^0 
&=& \partial_{\nu}E^{\nu}-\partial_k E^k \nonumber\\
&=& -{\textstyle\frac{4\pi}{c}}
\partial_{\nu}j^{\nu}-\partial_k E^k\,,
\label{eq:evol-constr}
\end{eqnarray}
where we used the identity (\ref{eq:identity-maxwell})
in the second step. Suppose now that on the initial 
surface of constant $x^0$ we put an electromagnetic field 
which satisfies the constraint, $E^0=0$, and which we evolve
according to $E^k=0$ (implying $\partial_kE^k=0$ on that 
initial surface). Then (\ref{eq:evol-constr}) shows that 
charge conservation is the necessary and sufficient condition
for the evolution to preserve the constraint. 

Let us return to GR now, where the overall situation is entirely 
analogous. Now we have four constraints
\begin{equation}
E^{0\nu}:= G^{0\nu}-\kappa T^{0\nu}=0
\label{eq:einst-constraints}
\end{equation}
and six evolution equations, which we write 
\begin{equation}
E^{ij}:= G^{ij}-\kappa T^{ij}=0\,.
\label{eq:einst-evol}
\end{equation}
The identity (\ref{eq:bianchi-id}) now implies 
\begin{equation}
\nabla_{\mu}E^{\mu\nu}=-\kappa\nabla_{\mu}T^{\mu\nu}\,,
\label{eq:identity-einstein}
\end{equation}
which parallels (\ref{eq:identity-maxwell}). 
Here, too, the time derivative of the constraints is 
easily calculated:
\begin{eqnarray}
\partial_0E^{0\nu}
&=&\nabla_{\mu}E^{\mu\nu}-\partial_kE^{k\nu}
    -\Gamma^0_{0\lambda}E^{\lambda\nu}
    -\Gamma^{\nu}_{0\lambda}E^{0\lambda}\nonumber\\
&=& -\kappa\nabla_{\mu}T^{\mu\nu}-\partial_kE^{k\nu}
    -\Gamma^0_{0\lambda}E^{\lambda\nu}
    -\Gamma^{\nu}_{0\lambda}E^{0\lambda}
\label{eq:evol-const-ein}
\end{eqnarray}
using (\ref{eq:identity-einstein}) in the last step. 
Now consider again the evolution of initial data from a 
surface of constant $x^0$. If they initially satisfy the 
constraints and are evolved via $E^{ij}=0$ (hence all 
spatial derivatives of $E^{\mu\nu}$ vanish initially) they 
continue to satisfy the constraints if and only if the 
energy momentum tensor of the matter satisfies
\begin{equation}
\nabla_{\mu}T^{\mu\nu}=0\,.
\label{eq:em-conservation}
\end{equation}
Hence we see that the `covariant conservation' of energy-momentum, 
expressed by (\ref{eq:em-conservation}), plays the same r\^ole in 
GR as charge conservation plays in EM. This means that you cannot 
just prescribe the motion of matter and then use Einstein's equations 
to calculate the gravitational field produced by that source. 
You have to move the matter in such a way that it satisfies 
(\ref{eq:em-conservation}). But note that at this point there is a 
crucial mathematical difference to charge conservation in EM:
charge conservation is a condition on the source \emph{only}, it 
does not involve the electromagnetic field. This means that you 
know \emph{a priori} what to do in order not to violate charge 
conservation. On the other hand, (\ref{eq:em-conservation}) involves 
the source \emph{and} the gravitational field. The latter enters 
through the covariant derivatives which involve the metric $g_{\mu\nu}$ 
through the connection coefficients (\ref{eq:christoffel}). 
Hence here (\ref{eq:em-conservation}) cannot be solved a priori 
by suitably restricting the motion of the source. Rather we have a 
consistency condition which \emph{mutually} links the problem of motion 
for the sources and the problem of field determination. 
It is this difference which makes the problem of motion in GR 
exceedingly difficult. (A brief and lucid 
presentation of this problem, drawing attention to its relevance 
in calculating the generation of gravitational radiation by 
self-gravitating systems, was given in \cite{Ehlers}.
A broader summary, including modern developments is \cite{Damour}).

For example, for pressureless dust represented by 
$T^{\mu\nu}=\rho c^2\, U^{\mu}U^{\nu}$, where $\rho$ is the local rest-mass
density and $U^{\mu}$ is the vector field of four-velocities of the 
continuously dispersed individual dust grains, (\ref{eq:em-conservation})
is equivalent to the two equations 
\begin{eqnarray}
\nabla_{\mu}(\rho U^{\mu})  &=&0\,,
\label{eq:restmass-conservation}\\
U^{\nu}\nabla_{\nu} U^{\mu} &=&0\,. 
\label{eq:geod-motion}
\end{eqnarray}
The first states the conservation of rest-mass. The second is equivalent 
to the statement that the vector field $U^{\mu}$ is geodesic, which means 
that its integral lines (the worldlines of the dust grains) are 
geodesic curves (\ref{eq:geodesic}) 
\emph{with respect to the metric $g_{\mu\nu}$}. Hence we see that in this 
case the motion of matter is fully determined by 
(\ref{eq:em-conservation}), i.e. by Einstein's equations, which imply 
(\ref{eq:em-conservation}) as integrability condition. 
This clearly demonstrates how the problem of motion is inseparably 
linked with the problem of field determination, and that these 
problems can only be solved simultaneously. The methods used today
use clever approximation schemes. For example, one can make use 
of the fact that there is a difference of one power in $\kappa$ between 
the field equations and their integrability condition. Hence, in an 
approximation in $\kappa$ it is consistent for the $n$th order 
approximation of the field equations to have the integrability 
conditions (equations of motions) satisfied to $n-1$st order.

Clearly the problem just discussed does not arise for the matter-free 
Einstein equations for which $T_{\mu\nu}\equiv 0$. Now recall that 
black holes too are described by the matter free equations. Hence the
mathematical problem just described does not occur in the discussion 
of their dynamics. In \emph{this} aspect the discussion of black hole 
scattering is considerably easier than e.g. that of neutron stars.

\section{The 3+1- split and the Cauchy initial-value problem}
We saw that the ten Einstein equations decompose into two sets 
of four and six equations respectively, four constraints which 
the initial data have to satisfy, and six equations driving the 
evolution. As a consequence there will be four dynamically 
undetermined components among the ten components of the 
gravitational field $g_{\mu\nu}$. The task is to parametrise the 
$g_{\mu\nu}$ in such a way that four dynamically undetermined 
functions can be cleanly separated from the other six. One way to 
achieve this is via the splitting of spacetime into space and time
(see \cite{Giulini 1998a} for a more detailed discussion). 
The four dynamically undetermined quantities will be the famous  
{\it lapse} (one function $\alpha$) and {\it shift} (three functions 
$\beta^i$). The dynamically determined quantity is the Riemannian 
metric $h_{ij}$ on the spatial 3-manifolds of constant time. These 
together parametrise $g_{\mu\nu}$ as follows:
\begin{equation}
ds^2=-\alpha^2\, (dx^0)^2+h_{ik}\,
   (dx^i+\beta^idx^0)(dx^k+\beta^kdx^0)\,.
\label{eq:split-metric}
\end{equation}
The physical interpretation of $\alpha$ and $\beta^i$ is as follows:
think of spacetime as the history of space. Each `moment' of time,
$x^0=t$, corresponds to an entire 3-dimensional slice $\Sigma_t$.
Obviously there is plenty of freedom how to 
`waft' space through spacetime. This freedom corresponds precisely to
the freedom to choose the 1+3 functions $\alpha$ and $\beta^i$. 
For one thing, you may freely specify how far for each parameter 
step $dt$ you push space in a perpendicular direction forward in time. 
This is controlled by $\alpha$, which is just the ratio $ds/dt$ of 
the \emph{proper} perpendicular distance between the hypersurfaces 
$\Sigma_t$ and $\Sigma_{t+dt}$. This speed may be chosen in a space 
and time dependent fashion, which makes $\alpha$ a function on 
spacetime. Second, given a point with coordinates $x^i$ on $\Sigma_t$. 
Going from $x^i$ in a perpendicular direction you meet $\Sigma_{t+dt}$ 
in a point with coordinates $x^i+dxi$, where $dx^i$ can be chosen 
at will. This freedom of moving the coordinate system around while 
evolving is captured by $\beta^i$; one writes $dx^i= \beta^idt$.
Clearly this moving around of the spatial coordinates can also be 
made in a space and time dependent fashion, so that the $\beta^i$ 
are functions of spacetime, too.

Let $n^{\mu}$ be the vector field in spacetime which is normal to 
the spatial sections of constant time. It is given by 
$n=\frac{1}{\alpha}(\partial/\partial x^0-\beta^i\partial/\partial x^i)$, 
as  one may  readily verify using (\ref{eq:split-metric}) 
(you have to check that $n$ is normalized and satisfies 
$g(n,\partial/\partial x^i)=0$). We define the 
\emph{extrinsic curvature}, $K_{ij}$, to be one-half the 
Lie derivative of the spatial metric in the direction of the 
normal:
\begin{equation}
K_{ij}:=\shalf L_nh_{ij}=
\frac{1}{2\alpha}\left(\frac{\partial h_{ij}}{\partial x^0}
-2D_{(i}\beta_{j)}\right)\,,
\label{eq:def-ext-curv}
\end{equation}
where $D$ is the spatial covariant derivative with respect to the 
metric $h_{ij}$. As usual, a round bracket around indices denotes 
their symmetrization. Note that, by definition, $K_{ij}$ is symmetric.
Finally we denote the Ricci scalar of $h_{ij}$ by $R^{\sss (3)}$. 

We can now write down the four constraints of the vacuum Einstein 
equations in terms of these variables:
\begin{eqnarray}
0&=&G(n,n)=\shalf(R^{\sss (3)}+K^{ij}K_{ij}-(K^i_i)^2)\,,
\label{eq:ham-constraint}\\
0&=&G(n,\partial/\partial x^j)=D_i(K^i_j-\delta^i_jK^k_k)\,.
\label{eq:mom-constraint}
\end{eqnarray}
(\ref{eq:ham-constraint}) and (\ref{eq:mom-constraint}) are 
referred to as \emph{Hamiltonian constraint} and \emph{momentum constraint}
respectively. The six evolution equations of second order in the time 
derivative can now be written as twelve equations of first order. 
Six of them are just (\ref{eq:def-ext-curv}), read as equation that
relates the time derivative $\partial h_{ij}/\partial x^0$ to the 
`canonical data' $(h_{ij},K_{ij})$. The other six equations, whose 
explicit form needs not concern us here (see e.g. \cite{Giulini 1998a}),  
express the time derivative of $K_{ij}$ in terms of the canonical data. 
Both sets of evolution equations contain on their right hand sides 
the lapse and shift functions, whose evolution is not determined
but must be specified by hand. This specification is a choice of 
gauge, without which one cannot determine the evolution of the 
physical variables $(h_{ij},K_{ij})$. 

The initial-data problem takes now the following form 
\begin{enumerate}
\item[I.]
Choose a topological 3-manifold $\Sigma$.
\item[II.]
Find on $\Sigma$ a Riemannian metric $h_{ij}$ and a symmetric
tensor field $K_{ij}$ which satisfy the constraints 
(\ref{eq:ham-constraint}) and (\ref{eq:mom-constraint}).
\item[III.]
Choose a lapse function $\alpha$ and a shift vector field 
$\beta^i$, both as functions of space and time, possibly according 
to some convenient prescription (e.g. singularity avoiding gauges, 
like maximal slicing).
\item[IV.]
Evolve initial data with these choices of $\alpha$ and $\beta^i$ 
according to the twelve equations of first order. By consistency 
of Einstein's equations the constraints will be preserved during 
this evolution, independent of the choices for $\alpha$ and $\beta^i$.
\end{enumerate}
The backbone of this setup is a mathematical theorem, which states 
that for any set of initial data, taken from a suitable function 
space, there is, up to diffeomorphism, a unique maximal Einstein 
spacetime developing from these data~\cite{Choquet-Geroch 1969}.

\section{Black hole data}
\subsection{Horizons}
By black hole data we understand vacuum data which contain 
\emph{apparent horizons}. The informal definition of an apparent 
horizon is that it is the boundary of a trapped region, which 
means that its orthogonal outgoing null rays must have zero 
divergence. (Inside the trapped region they converge for any 
2-surface, by definition of `trapped region'.) 
The Penrose-Hawking singularity theorems state that the existence 
of an apparent horizon implies that the evolving spacetime will 
be singular (assuming the strong energy-condition). Given also 
the condition that singularities cannot be seen by observers far off,
a condition usually called \emph{cosmic censorship}, one infers the 
existence of an \emph{event horizon} and hence a black hole. 
One can then show that the intersection of the event horizon with 
the spatial hypersurface lies on or outside the apparent 
horizon (for stationary spacetimes they coincide). 
The reason why one does not deal with event horizons directly is 
that you cannot tell whether there exists one by just looking at 
initial data. In principle you would have to evolve them to the 
infinite future, which is beyond our abilities in general. In contrast,
apparent horizons can be recognized once the data on an initial slice 
are given. The formal definition of an apparent horizon is the 
following: given initial data $(\Sigma,h_{ij},K_{ij})$ and an embedded 
2-surface $\sigma\subset\Sigma$ with outward pointing normal $\nu^i$. 
$\sigma$ is an apparent horizon if and only if the following relation 
between $K_{ij}$, the extrinsic curvature of $\Sigma$ in spacetime, 
and $k_{ij}$, the extrinsic curvature of $\sigma$ in $\Sigma$, is 
satisfied:
\begin{equation}
q^{ij}k_{ij}=-q^{ij}K_{ij}\,,
\label{eq.def-apparent-horizon}
\end{equation}
where $q_{ij}:=h_{ij}-\nu_i\nu_j$ is just the induced Riemannian 
metric on $\sigma$, so that (\ref{eq.def-apparent-horizon}) simply 
says that the restriction of $K_{ij}$ to the tangent space of $\sigma$ 
has opposite trace to $k_{ij}$. (The minus sign on the rhs of 
(\ref{eq.def-apparent-horizon}) signifies a \emph{future} apparent 
horizon corresponding to a \emph{black} hole which has a \emph{future} 
event horizon. A plus sign would signify a past apparent horizon 
corresponding to a `\emph{white} hole' with \emph{past} event horizon.)
This means that once we have the data $(\Sigma,h_{ij},K_{ij})$ we 
can in principle find all 2-surfaces $\sigma\subset\Sigma$ for 
which (\ref{eq.def-apparent-horizon}) holds and therefore find 
all apparent horizons.

\subsection{Poincar\'e charges}
By Poincar\'e charges we shall understand quantities like 
mass, linear momentum, and angular momentum. In GR they are 
associated to an asymptotic Poincar\'e symmetry 
(see \cite{Beig-Murchadha}), provided that the data 
$(\Sigma,h_{ij},K_{ij})$ are \emph{asymptotically flat} in a 
suitable sense, which we now explain. Topologically asymptotic flatness
means that the non-trivial `topological features' of $\Sigma$ should 
all reside in a bounded region and not `pile up' at infinity. More 
formally this is expressed by saying that there is a bounded region 
$B\subset\Sigma$ such that $\Sigma-B$ (the complement of $B$) consists 
of a finite number of disjoint pieces, each of which looks topologically 
like the complement of a ball in $\reals^3$. These asymptotic pieces 
are also called the \emph{ends} of the manifold $\Sigma$. Next comes the 
geometric restriction imposed by the condition of asymptotic flatness.
It says that for each end there is an asymptotically euclidean 
coordinate-system $\{x^1,x^2,x^3\}$ in which the fields $(h_{ij},K_{ij})$ 
have the following fall-off for $r\rightarrow\infty$ 
($r=\sqrt{(x^1)^2+(x^2)^2+(x^3)^2}$, $n^k=x^k/r$)
\begin{eqnarray}   
h_{ij}(x^k)&=&\delta_{ij}+\frac{s_{ij}(n^k)}{r}
+O(r^{-1-\epsilon})\,,
\label{eq:h-falloff}\\
K_{ij}(x^k)&=&\frac{t_{ij}(n^k)}{r^2}
+O(r^{-2-\epsilon})\,.
\label{eq:K-falloff}
\end{eqnarray}
Moreover, in order to have convergent expressions for physically 
relevant quantities, like e.g. angular momentum (see below), the 
field $s_{ij}$ must be an \emph{even} function of its argument, 
i.e. $s_{ij}(-n^k)=s_{ij}(n^k)$, and $t_{ij}$ must be an \emph{odd} 
function, i.e.  $t_{ij}(-n^k)=-t_{ij}(n^k)$. 

Under these conditions each end can be assigned mass, momentum, and 
angular momentum, which are conserved during time evolution. 
They may be computed by integrals over 2-spheres in the 
limit the spheres are pushed to larger and larger radii into the 
asymptotically flat region of that end. These so-called ADM-Integrals 
(first considered by Arnowitt, Deser, and Misner in \cite{ADM}) are 
given by the following expressions, which we give in `geometric' units
(meaning that in order to get them in standard units one has to multiply 
the mass expression given below by $1/\kappa$ and the linear and angular 
momentum by $c/\kappa$):
\begin{eqnarray}
M&=&\lim_{r\rightarrow\infty}
    \int_{S^2(r)}\delta^{ij}
    (\partial_ih_{jk}-\partial_kh_{ij})n^k\,d\sigma
    \label{eq:ADM-mass}\\
P^i&=&\lim_{r\rightarrow\infty}
    \int_{S^2(r)} (K^i_k-\delta^i_kK^j_j)n^k\,d\sigma
    \label{eq:ADM-momentum}\\
S^i&=&\lim_{r\rightarrow\infty}
      \int_{S^2(r)}\varepsilon_{ijl}x^j(K^l_k-\delta^l_kK^n_n)
    n^k\, d\sigma
\label{eq:ADM-angular-momentum}
\end{eqnarray}

\subsection{Maximal and time-symmetric data}
The constraints (\ref{eq:ham-constraint},
\ref{eq:mom-constraint}) are too complicated to be solved 
in general. Further conditions are usually imposed to reduce the 
complexity of the problem: data $(h_{ij},K_{ij})$ are called 
\emph{maximal} if $K^i_i=h^{ij}K_{ij}=0$. The name derives from 
the fact that $K^i_i=0$ is the necessary and sufficient condition 
for a hypersurface to have stationary volume to first order with 
respect to deformations in the ambient spacetime. Even though 
stationarity does generally not imply extremality one calls such 
hypersurfaces maximal. Note also that since spacetime is a 
Lorentzian manifold, extremal spacelike hypersurfaces will be of 
maximal rather than minimal volume. In contrast, in Riemannian 
manifolds one would speak of minimal surfaces. 

A much stronger condition is to impose $K_{ij}=0$, which as seen 
from (\ref{eq:ADM-momentum}) and (\ref{eq:ADM-angular-momentum}) 
implies that all momenta and angular momenta vanish. Only the mass 
is now allowed to be non-zero. Such data are called 
\emph{time-symmetric} since for them $h_{ij}$ is momentarily static
as seen from (\ref{eq:def-ext-curv}). This implies that the 
evolution of such data into the future and into the past will coincide
so that the developed spacetime will have a time-reversal symmetry 
which pointwise fixes the initial surface where $K_{ij}=0$. This 
surface is therefore also called the \emph{moment of time-symmetry}. 
Time-symmetric data can still represent configurations of any number 
of black holes without angular momenta which are momentarily at rest. 
Note also that for time-symmetric data the condition 
(\ref{eq.def-apparent-horizon}) for an apparent horizon is equivalent 
to the tracelessness of the extrinsic curvature of $\sigma$. 
\emph{Hence for time symmetric data apparent horizons are minimal surfaces.} 

We add one more general comment concerning submanifolds. A vanishing 
extrinsic curvature is equivalent to the property that each 
geodesic of the ambient space, which starts on, and tangent to, the 
submanifold, will always run entirely inside the submanifold. 
Therefore, submanifolds with vanishing extrinsic curvature are 
called \emph{totally geodesic}. Now, if the ambient space allows for 
an isometry (symmetry of the metric), whose fixed-point set is the 
submanifold in question, like for the time-reversal transformation 
just discussed, the submanifold must necessarily be totally geodesic. 
To see this, consider a geodesic of the ambient space which 
starts on, and tangent to, the submanifold. Assume that this geodesic 
eventually leaves the submanifold. Then its image under the isometry 
would again be a geodesic (since isometries always map geodesics to 
geodesics) which is different from the one we started from. But this 
is impossible since they share the same initial conditions which are 
known to determine the geodesic uniquely. Hence the geodesic cannot 
leave the submanifold, which proves the claim. We will later have 
more opportunities to identify totally geodesic submanifolds---namely 
apparent horizons---by their property of being fixed-point sets of 
isometries.

\subsection{Solution strategy for maximal data}
Possibly the most popular approach to solving the constraints 
is the \emph{conformal technique} due to York, Lichnerowicz and
others (see \cite{York 1979} for a review). The basic idea is to 
regard the Hamiltonian constraint (\ref{eq:ham-constraint}) as 
equation for the conformal factor of the metric $h_{ij}$ and 
\emph{freely} specify the complementary information, called the 
conformal equivalence-class of $h_{ij}$. More concretely, 
this works as follows:
\begin{enumerate}
\item[I.]
Choose unphysical (`hatted') quantities $(\hat{h}_{ij},\hat{K}_{ij})$,
where $\hat{h}_{ij}$ is a Riemannian metric on $\Sigma$ and 
$\hat{K}_{ij}$ is symmetric, trace and divergence free:
\begin{equation}
\hat{h}^{ij}\hat{K}_{ij}=0,\qquad \hat{D}^i\hat{K}_{ij}=0\,,
\label{eq:hat-constraints}
\end{equation}
where $\hat{D}$ is the covariant derivative with respect to 
$\hat{h}_{ij}$.
\item[II.]
Solve the (quasilinear elliptic) equation for a positive, real valued 
function $\Phi$ with boundary condition 
$\Phi(r\rightarrow\infty)\rightarrow 1$, where  
$\hat{\Delta}=\hat{h}^{ij}\hat{D}_i\hat{D}_j$:
\begin{equation}
\hat{\Delta}\Phi+\eights \hat{K}^{ij}\hat{K}_{ij}\Phi^{-7}=0
\label{eq:Lichnerowicz}
\end{equation}
\item[III.]
Using the solution of (\ref{eq:Lichnerowicz}), define physical 
(`unhatted') quantities by
\begin{eqnarray}
h_{ij}&=&\Phi^4\hat{h}_{ij}\,,
\label{eq:h-solution}\\
K_{ij}&=&\Phi^{-2}\hat{K}_{ij}\,.
\label{eq:K-solution}
\end{eqnarray}
\emph{These will satisfy the constraints (\ref{eq:ham-constraint},
\ref{eq:mom-constraint})!}
\end{enumerate}

\subsection{Explicit time symmetric data}
Before we say a little more about maximal data we wish to present 
some of the most popular examples for time symmetric data some of 
which are also extensively used in numerical simulations. Hopefully 
these examples let you gain some intuition in the geometries and 
topologies involved and also let you anticipate the richness that 
a variable space-structure gives to the solution space of one of the 
simplest equations in physics: the Laplace equation. 

Restricting the solution strategy, outlined above, to the time-symmetric 
case one first observes that for $K_{ij}=0$ one has $\hat{K}_{ij}=0$. 
The momentum constraint (\ref{eq:mom-constraint}) is automatically 
satisfied and all that remains is equation (\ref{eq:Lichnerowicz}), 
which now simply becomes the Laplace equation for the single scalar 
function $\Phi$ on the Riemannian manifold $(\Sigma,\hat{h}_{ij})$. 

We now make a further simplifying assumption, namely that $\hat{h}_{ij}$
is in fact the \emph{flat} metric. This will restrict our solution
$h_{ij}$ to a \emph{conformally flat} geometry. It is not obvious 
how severe the loss of physically interesting solutions is by restricting 
to conformally flat metrics. But we will see that the latter already 
contain many interesting and relevant examples. 

So let us solve Laplace's equation in flat space! Remember that $\Phi$
must be positive and approach 1 at spatial infinity (asymptotic
flatness). We cannot take $\Sigma=\reals^3$ since the only solution 
to the Laplace equation in $\reals^3$ which asymptotically approaches 
1 is identically 1. We must allow $\Phi$ to blow up at some points, 
which we can then remove from the manifold. In this way we let 
the solution tell us what topology to choose in order to have an 
everywhere regular solution. You might think that just removing 
singular points would be rather cheating, since the resulting manifold 
may turn out to be incomplete, that is, can be hit by a curve after 
finite proper length (you can go `there'), even though $\Phi$ and hence 
the physical metric blows up at this point. If this were the case one 
definitely had to say what a solution on the completion would be. 
But, as a matter of fact, this cannot happen and the punctured space 
will turn out to be complete in the physical metric.

\subsubsection{One black hole}
The simplest solution with one puncture (at $r=0$) is just
\begin{equation}
\Phi(r,\theta,\varphi) =1+\frac{a}{r}\,,
\label{eq:Phi-one-hole}
\end{equation}
where $a$ is a constant which we soon interpret and which must be 
positive in order for $\Phi$ to be positive everywhere. 
We cannot have other multipole contributions since they inevitably 
would force $\Phi$ to be negative somewhere. What is the geometry of 
this solution? The physical metric is 
\begin{equation}
ds^2=\left(1+\frac{a}{r} \right)^4
(dr^2+r^2\,d\theta^2+\sin^2\theta\,d\varphi^2)
\label{eq:metric-one-hole}
\end{equation}
which is easily checked to be invariant under the inversion-transformation
on the sphere $r=a$:
\begin{equation}
r       \rightarrow \frac{a^2}{r},\quad 
\theta  \rightarrow \theta,           \quad
\varphi \rightarrow \varphi\,.
\label{eq:inversion-trans}
\end{equation}
This means that the region $r>a$ just looks like the region 
$r<a$ and that the sphere $r=a$ has the smallest area among 
all spheres of constant radius. It is a minimal surface, in 
fact even a totally geodesic submanifold, since it is the 
fixed point set of the isometry (\ref{eq:inversion-trans}). 
Hence it is an apparent horizon, whose area follows from 
(\ref{eq:metric-one-hole}): 
\begin{equation}
A=16\pi (2a)^2\,.
\label{eq:area-horizon}
\end{equation}
Our manifold thus corresponds to a black hole. Its mass can easily 
be computed from (\ref{eq:ADM-mass}); one finds $m=2a$. 
\begin{figure}
\noindent
\centering\epsfig{figure=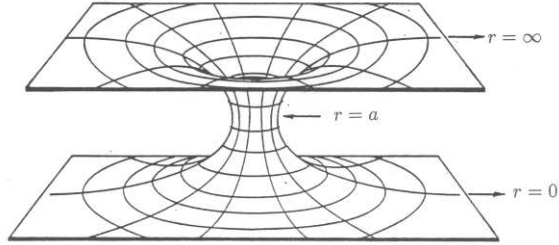, width=0.6\linewidth}
\caption{\small One black hole}
\label{fig:1BH-Schw}
\end{figure}
This manifold has two ends, one for $r\rightarrow\infty$ and 
one for $r\rightarrow 0$. They have the same geometry and hence 
the same ADM mass, as it must be the case since individual and 
total mass clearly coincide for a single hole.

The data just written down correspond to the `middle' slice 
right across the  Kruskal (maximally extended Schwarzschild) 
manifold. Also, (\ref{eq:metric-one-hole}) is just the spatial 
part of the Schwarzschild metric in isotropic coordinates. 
Hence we know its entire future development in analytic form. 
Already for two holes this is not the case anymore. Even the 
simplest two body problem---head-on collision--- has not been 
solved analytically in GR.

\subsubsection{Two black holes} 
\label{subsubsec:two-BH}
\begin{figure}
\noindent
\centering\epsfig{figure=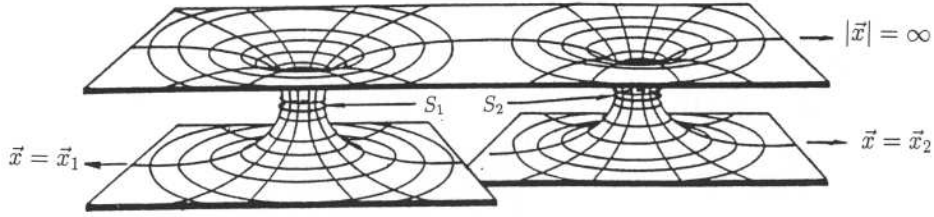, width=\linewidth}
\caption{\small Two black holes well separated}
\label{fig:2BH-apart}
\end{figure}

There is an obvious generalization of (\ref{eq:Phi-one-hole}) by 
allowing two `monopoles' of strength $a_1$ and $a_2$ at the punctures 
$\vec x=\vec x_1$ and $\vec x=\vec x_2$ respectively. The 3-metric then 
reads

\begin{equation}
ds^2=\left(1+\frac{a_1}{\vert\vec x-\vec x_1\vert}
            +\frac{a_2}{\vert\vec x-\vec x_2\vert}\right)^4
(dr^2+r^2\,d\theta^2+\sin^2\theta\,d\varphi^2). 
\label{eq:metric-two-hole-Schw}
\end{equation}
The manifold has now three asymptotically flat ends, one for 
$\vert\vec x\vert\rightarrow\infty$, where the overall ADM mass $M$ 
is measured, and one each for $\vert\vec x-\vec x_{1,2}\vert\rightarrow 0$.
To see the latter, it is best to write the metric 
(\ref{eq:metric-two-hole-Schw}) in spherical polar coordinates 
$(r_1,\theta_1,\varphi_1)$ centered at $\vec x_1$, and then introduce 
the inverted radial coordinate given by $\bar{r}_1=a_1^2/r_1$. In the 
limit $\bar{r}_1\rightarrow\infty$ the metric then takes the form 
\begin{equation}
ds^2=\left(1+\frac{a_1(1+a_2/r_{12})}{\bar{r}_1}+O((1/\bar{r}_1)^{2})
\right)^4
(d\bar{r}^2+\bar{r}^2\,(d\theta_1^2+\sin^2\theta_1 d\varphi_1^2))
\label{eq:2BH-asymp-metric}
\end{equation}
where $r_{12}=\vert\vec x_1-\vec x_2\vert$. This looks just like a 
one-hole metric (\ref{eq:metric-one-hole}). Hence, if the 
black holes are well separated (compared to their size),  the 
two-hole geometry looks like that depicted in 
figure~\ref{fig:2BH-apart}.
By comparison with the one-hole metric we can immediately write down 
the ADM masses corresponding to the three ends 
$r,\bar{r}_{1,2}\rightarrow\infty$ respectively:
\begin{eqnarray}
M=2(a_1+a_2),\quad 
m_{1,2}=2a_{1,2}(1+\chi_{1,2}),
\quad\mbox{where}\quad 
\chi_{1,2}=\frac{a_{2,1}}{r_{12}}\,.\qquad
\label{eq:2BH-masses}
\end{eqnarray}
Momenta and angular momenta clearly vanish (moment of time symmetry). 
Still assuming well separated holes, i.e. $\chi_i=a_{i}/r_{12}\ll 1$, 
we can calculate the binding energy $\Delta E=M-m_1-m_2$ as 
function of the masses $m_{i}$ and $r_{12}$ and get 
\begin{eqnarray}  
\Delta E=-\frac{m_1m_2}{r_{12}}
\left(1-\frac{m_1+m_2}{2r_{12}} + O((m_{1,2}/r_{12})^2)\right)\,.
\label{eq:2BH-binding-energy}
\end{eqnarray}
The leading order is just the Newtonian expression for the binding 
energy of two point-particles with masses $m_{1,2}$ at distance 
$r_{12}$. But there are corrections to this Newtonian form which tend 
to diminish the Newtonian value. Note also that 
(\ref{eq:2BH-binding-energy}) is still not in a good form since 
$r_{12}$ is not an invariantly defined geometric distance measure. 
As such one might use the length $\ell$ of the shortest geodesic 
joining the two apparent horizons $S_1$ and $S_2$. Unfortunately these 
horizons are not easy to locate analytically and hence no closed form 
of $\ell(m_1,m_2,r_{12})$ exists which could be inverted to eliminate 
$r_{12}$ in favour of $\ell$.

Due to the difficulty to analytically locate the two apparent horizons
we also cannot write down an analytic expression for their area. But we 
can give upper and lower bounds as follows:
\begin{equation} 
16\pi (2a_i)^2 < A_i < 16\pi [(2a_i(1+\chi_i)]^2=16\pi m_i^2\,.
\label{eq:area-bounds}
\end{equation}
The lower bound simply follows from the fact that the two-hole
metric (\ref{eq:metric-two-hole-Schw}), if written down in terms of 
spherical polar coordinates about any of its punctures, equals the 
one hole metric (\ref{eq:metric-one-hole}) plus a positive definite
correction. The upper bound follows from the so-called Penrose 
inequality in Riemannian Geometry (proven in \cite{Huisken and 
Ilmanen}), which directly states that $16\pi m^2\geq A$ for each 
asymptotically flat end, where $m$ is the mass according to 
(\ref{eq:ADM-mass}) and $A$ is the area of the outermost (as seen 
from that end) minimal surface.  

\begin{figure}
\noindent
\centering\epsfig{figure=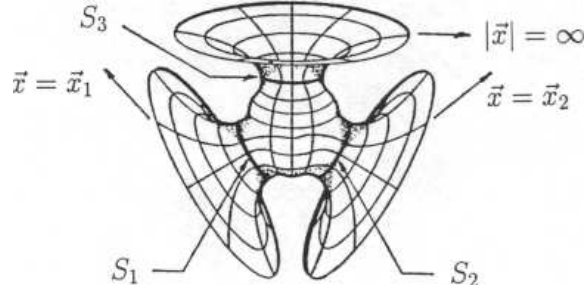, width=0.6\linewidth}
\caption{\small Two black holes after merging}
\label{fig:2BH-close}
\end{figure}

If the two holes approach each other to a distance comparable to the sizes 
of the holes the geometry changes in an essential way. This is shown in 
figure~\ref{fig:2BH-close}.
The most important new feature is that new minimal surfaces form, 
in fact two \cite{Cadez}, which both enclose the two holes. 
The outermost of these, as seen from the upper end, denoted by $S_3$ 
in figure \ref{fig:2BH-close}, corresponds to the apparent horizon 
of the newly formed `compound' black hole which contains the two old
ones.  For two black holes of equal mass, i.e. $a_1=a_2=a$,  this happens 
approximately for a parameter ratio of $a/r_{12}=0.65$, which in an 
approximate numerical translation into the ratio of individual hole 
mass to geodesic separation reads $m/\ell\approx 0.26$.

\subsubsection{ More than two black holes}
The method can be generalized in a straightforward manner to 
any number $n$ of black holes with parameters 
$(a_i,\vec x_i)$, 
$i=1,\cdots n$, for the punctures. The manifold $\Sigma$ has 
now $n+1$ ends, one for $\vert\vec x\vert\rightarrow\infty$ and 
one for each $\vec x\rightarrow\vec x_i$. The expressions for 
the metric and masses are then given by the obvious generalizations 
of (\ref{eq:metric-two-hole-Schw}) and (\ref{eq:2BH-masses}) 
respectively.

\subsubsection{Energy bounds from Hawking's area law}
Loosely speaking, Hawking's area law states that the surface of a 
black hole cannot decrease with time. (See \cite{Giulini 1998b}
for a simple and complete outline of the traditional and technically 
slightly restricted  version and \cite{Chrusciel} for the technically
most complete proof known today.) Let us briefly explain this statement. 
If $\Sigma$ is a Cauchy surface (a spacelike hypersurface in spacetime) 
and $\mathcal{H}$ the event horizon (a lightlike hypersurface in spacetime),
the two intersect in a number of components (spacelike 2-manifolds), each 
of which we assume to be a two-sphere. Each such two-sphere is called 
the surface of a black hole at time $\Sigma$. Let us pick one of them and 
call it $B$. Consider next a second Cauchy surface $\Sigma'$ which lies 
to the future of $\Sigma$. The outgoing null rays of $B$ intersect 
$\Sigma'$ in a surface $B'$, and the statement is now that the area
of $B'$ is larger than or equal to the area of $B$ (to prove this one 
must assume the strong energy condition). Note that we deliberately 
left open the possibility that $B'$ might be a proper subset of a 
black hole surface at time $\Sigma'$, namely in case the original 
hole has merged in the meantime with another one. If this does not 
happen $B'$ may be called the surface of the \emph{same} black hole at 
the later time $\Sigma$. 

Following an idea of Hawking's \cite{Hawking 1971}, this can be applied 
to the future evolution of multi black hole data as follows. As we 
already mentioned, the event horizon lies 
on or outside the apparent horizon. Hence the area of the `surface' 
(as just defined) of a black hole is bounded below by the area of the 
corresponding apparent horizon, which in turn has the lower bound 
stated in (\ref{eq:area-bounds}). Suppose that after a long time our 
configuration settles in an approximately stationary state, at least 
for some interior region where no gravitational radiation is emitted 
anymore. Since our data have zero linear and angular momentum, the 
final state is static and uniquely given by a single Schwarzschild 
hole of some final mass $M_{\f}$ and corresponding surface area 
$A_{\f}=16\pi M_{\f}^2$. This is a direct consequence of known black hole 
uniqueness theorems (see \cite{Heusler} or p.~157-186 of \cite{Hehl} 
for a summary). By the area theorem $A_{\f}$ is not less than the sum 
of all initial apparent horizon surface areas. This immediately gives
\begin{equation}
M_{\f}\geq\sqrt{\sum_i A_i^{\i}/16\pi}\geq 2\sqrt{\sum_i^n a_i^2}\,.
\label{eq:mass estimate}
\end{equation} 
In passing we remark that applied to a single black hole this argument 
shows that it cannot lose its mass below the value 
$m_{\sss\rm ir}:=\sqrt{A^{\i}/16\pi}$, called its \emph{irreducible mass}.
Back to the multi-hole case, the total initial mass is given by the 
straightforward generalization of 
(\ref{eq:2BH-masses}): 
\begin{equation}
M_{\i}=\sum_i m_i=2\sum_i a_i(1+\chi_i),\quad
\mbox{where}\quad
\chi_i=\sum_{i\not=k} 
\frac{a_k}{\vert\vec x_i-\vec x_k\vert}\,. 
\label{eq:mass-init}
\end{equation}
Using these two equations we can write down a lower bound for the 
fractional energy loss into gravitational radiation:
\begin{equation}
\frac{\Delta M}{M}:=\frac{M_{\i}-M_{\f}}{M_{\i}}\leq 1-
\frac{\sqrt{\sum_i a_i^2}}{\sum_i a_i(1+\chi_i)}\,.
\label{eq:total-smash-efficiency1}
\end{equation}
For a collision of $n$ initially widely separated ($\chi_i\rightarrow 0$) 
holes of equal mass this becomes 
\begin{equation}
\frac{\Delta M}{M}=1-\sqrt{1/n}\,.
\label{eq:total-smash-efficiency2} 
\end{equation}
For just two holes this means that at most 29\% of their total rest mass
can be radiated away. But this efficiency can be enhanced if the energy 
is distributed over a larger number of black holes.

Another way to raise the upper bound for the efficiency is to consider 
spinning black holes. For two holes the maximal value of 50\% can be 
derived by starting with two extremal black holes (i.e. of maximal 
angular momentum: $J=m^2$ in geometric units) which merge to form 
a single unspinning black hole \cite{Hawking 1971}. 

One can also envisage a situation where one hole participates in 
a scattering process but does not merge. Rather it gets kicked out 
of the collision zone and settles without spin (for simplicity) 
in an quasistationary state (for some time) far apart. 
The question is what fraction of energy the area theorem allows it 
to lose. Let this be the $k$th hole. 
Then $m^{\f}_k\geq 2a_k=m^{\i}_k/(1+\chi_k)$. Hence
\begin{equation}
\frac{m^{\i}_k-m^{\f}_k}{m^{\i}_k}\leq \frac{\chi_k}{1+\chi_k}< 1\,,
\label{eq:single-smash-efficiency}
\end{equation}
showing that an appreciable efficiency can only be obtained if 
the data are such that $\chi_k$ is not too close to zero. 
This means that the $kth$ hole was originally not too far from 
the others. This seems an unlikely process. Hence it is difficult
to extract energy from a single unspinning hole. 

For a single spinning black hole the situation is again different. 
Spinning it down from an extreme state to zero angular momentum
sets an upper bound for the efficiency from the area law of 29\%.
This follows easily from the following relation between mass, 
irreducible mass, and angular momentum
for a Kerr black hole (see e.g. \cite{MTW}, formula (33.60)):
\begin{equation}
m^2=m_{\sss\rm ir}^2+J^2/4m_{\sss\rm ir}^2
\label{eq:Kerr-mass}
\end{equation}
Setting $J=m^2$ one solves for $m_{\sss\rm ir}/m=\sqrt{1/2}$; hence 
$(m-m_{\sss\rm ir})/m=1-\sqrt{1/2}\approx 0.29$. It can moreover be 
shown \cite{Christodoulou} that this limit can be 
(theoretically) realised by the Penrose process (compare Frolov's 
lecture).

Needless to say that realistic processes may have far less 
efficiency than this theoretical bounds from the area law alone 
indicate. Recent numerical studies of the head-on (i.e. zero
angular momentum) collision of two equal-mass black holes give a 
radiated energy in units of the total energy of only $10^{-3}$ 
\cite{Baker 2000}. With angular momentum the efficiency is of course 
expected to be much better. Here recent numerical investigations 
give an estimate of $3\times 10^{-2}$ for an inspiral of two equal-mass 
non-spinning black holes from the innermost stable circular orbit
\cite{Baker 2001}: still a long way from the theoretical upper bound.

\subsubsection{Other topologies}
Other topologies can be found which support initial data with apparent 
horizons. For example, instead of the `Schwarzschild' manifold with 
$n+1$ ends for $n$ black holes one can find one which has just two 
ends for any number $\geq 2$ of holes and which has been termed the 
Einstein-Rosen manifold \cite{Lindquist 1963}.
\begin{figure}
\noindent
\begin{minipage}[b]{0.45\linewidth}
\centering\epsfig{figure=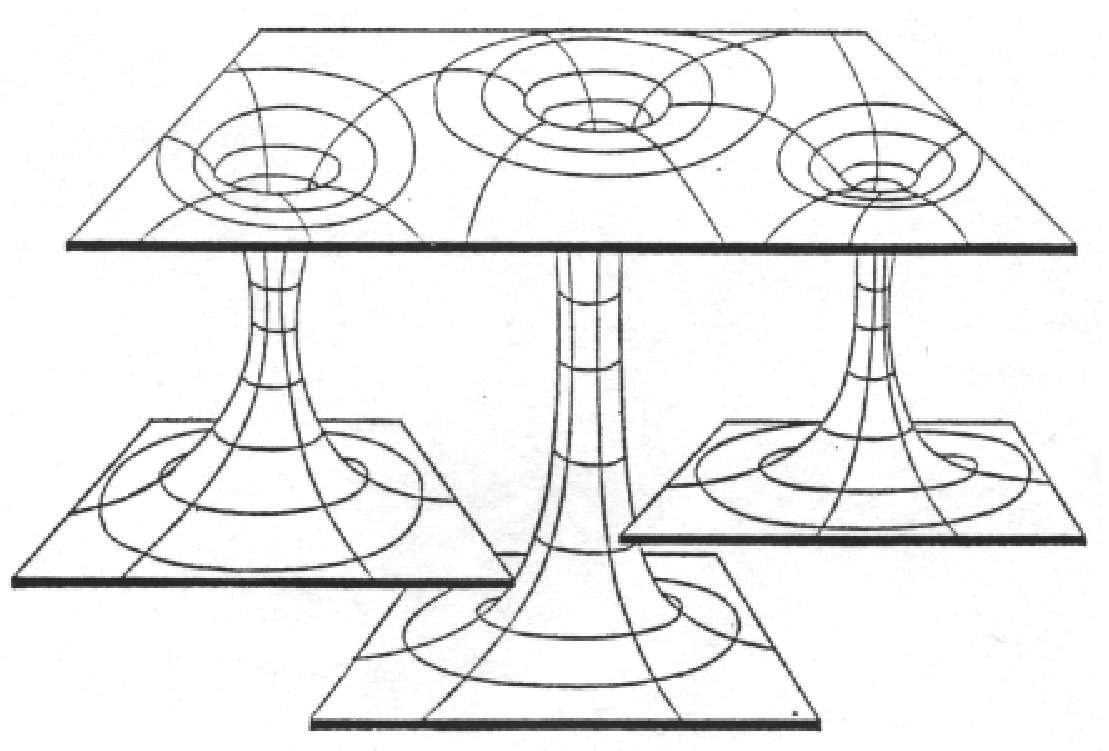, width=\linewidth}
\caption{\small Multi-Schwarzschild}
\label{fig:3BH-Schw}
\end{minipage}
\hfill
\begin{minipage}[b]{0.5\linewidth}
\centering\epsfig{figure=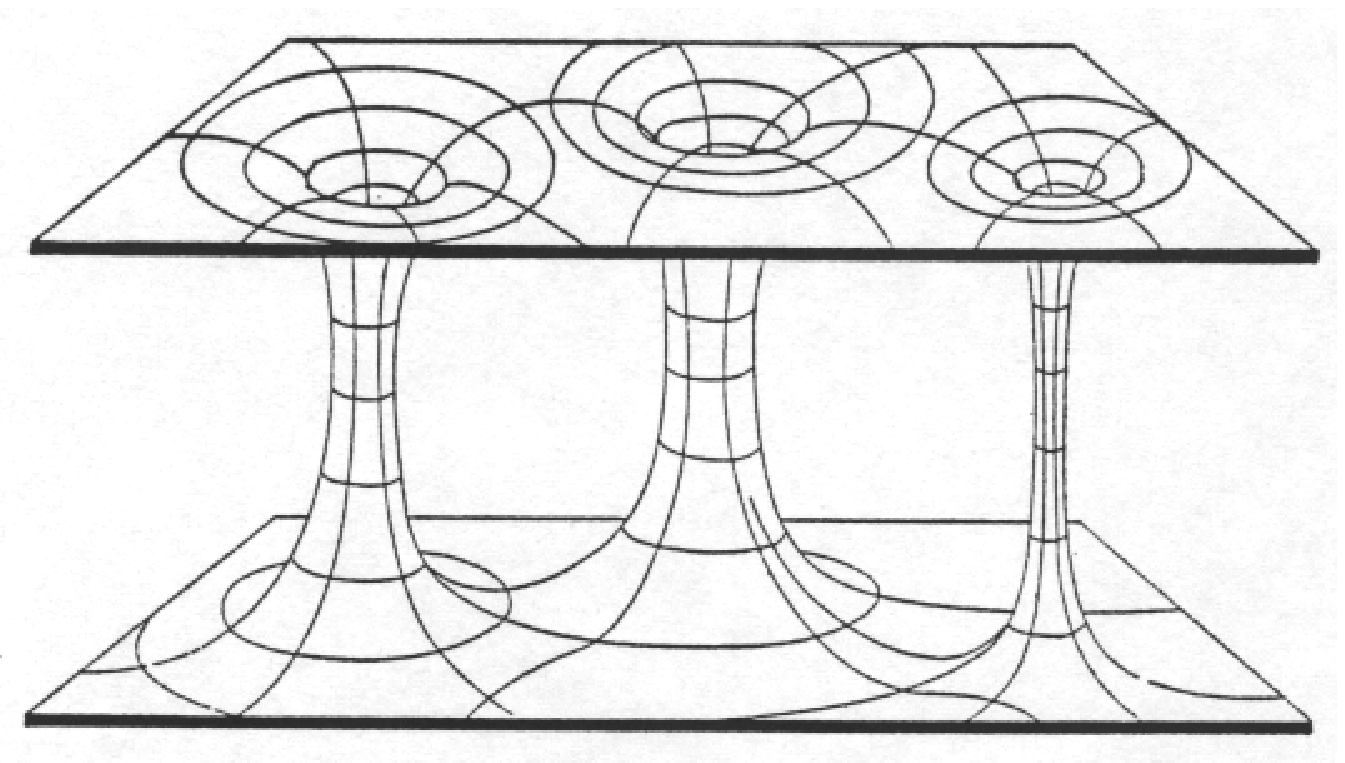, width=\linewidth}
\caption{\small Einstein-Rosen manifold}
\label{fig:3BH-ER}
\end{minipage}
\end{figure}
The difference to the data already discussed does not primarily lie in 
the physics they represent. After all their different topologies are 
hidden behind event horizons for the outside observer, even though their 
interaction energies are slightly different 
\cite{Giulini 1990}\cite{Giulini 1998a}. 
However, the point we wish to stress here is that such data can 
analytically and numerically be more convenient, despite the fact that 
the underlying manifold might seem \emph{topologically} more complicated. 
The reason is that these data have more symmetries, and that coordinate 
systems can be found for which these symmetries take simple analytic 
expressions. For example, in the Einstein-Rosen manifold the upper and 
lower ends are isometrically related by reflections about the minimal 
2-spheres in each connecting tube, with the fixed-point sets being the 
apparent horizons. Hence all the apparent horizons can analytically be 
easily located in the multi-hole Einstein-Rosen manifold, in contrast 
to the multi-hole Schwarzschild manifold.

\begin{wrapfigure}{r}{0.6\linewidth}
\centering\epsfig{figure=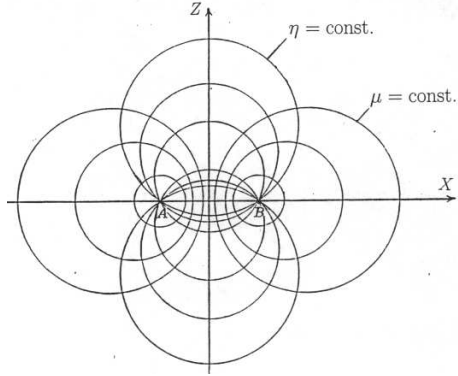, width=0.8\linewidth}
\caption{bipolar coordinates}
\end{wrapfigure}
Let us briefly explain this for two holes of equal mass. Here one starts 
again from $\reals^3$ coordinatised by spherical bi-polar coordinates. 
These are obtained from bi-polar coordinates $(\mu,\eta)$ in the $xz$ plane 
by adding an azimuthal angle $\varphi$ corresponding to a rotation about 
the $z$ axis, just like ordinary spherical polar coordinates are produced 
from ordinary polar coordinates in the $xz$-plane. The coordinates 
$(\mu,\eta)$ parametrise the $xz$-plane according to 
$\exp(\mu-i\eta)=(\xi+c)/(\xi-c)$, where $\xi=z+ix$ and $c>0$ is a 
constant.  The lines of constant $\mu$ intersect those of constant 
$\eta$ orthogonally. Both families consist of circles; those in the first 
family are centered on the $z$-axis with radii $c/\sinh \mu$ at 
$z=c\,\coth \mu$, and those in the second family on the $x$-axis with 
radii $c/\vert\sin\eta\vert$ at $\vert x\vert=c\,\cot \eta$.

Following an idea of Misner \cite{Misner 1963}, one can borrow 
the method of images from electrostatics (see. e.g. chapter~2.1 in 
\cite{Jackson}) to construct solutions $\Phi$ to the Laplace equation 
such that the metric $h_{ij}=\Phi^4\delta_{ij}$ has a number of 
reflection isometries about two-spheres, one for each hole. 
In the two hole case one uses the two two-spheres $\mu=\pm\mu_0$
for some $\mu_0>0$, which then become the apparent horizons. Using 
these isometries we can take two copies of our initial manifold, 
excise the balls $\vert\mu\vert>\mu_0$ and glue the two remaining parts 
`back to back' along the two boundaries $\mu=\mu_0$ and $\mu=-\mu_0$.
The isometry-property is necessary so that the metric continues  to 
be smooth across the seam. This gives an Einstein-Rosen manifold with 
two tubes (or `bridges', as they are sometimes called) connecting two 
asymptotically flat regions. 

In fact, we could have just taken \emph{one} copy of the original manifold, 
excised the balls $\vert\mu\vert>\mu_0$, and \emph{mutually} glued together 
the two boundaries $\mu=\pm\mu_0$. This also gives a smooth metric  
across the seam and results in a manifold known as the Misner 
wormhole \cite{Misner 1960}. 
\begin{figure}
\noindent
\centering\epsfig{figure=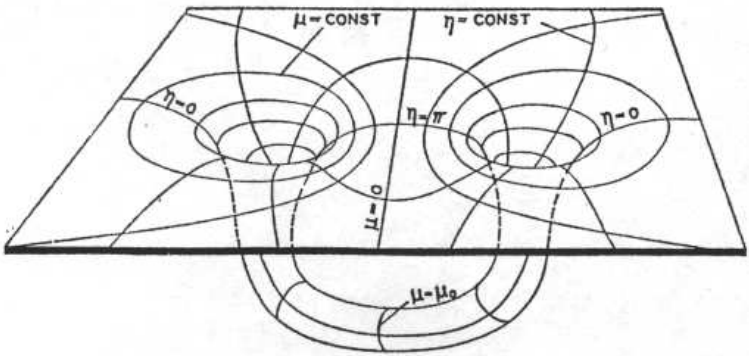, width=0.8\linewidth}
\caption{\small The Misner Wormhole representing two black holes}
\label{fig:1BH-Misner}
\end{figure}
Metrically the Misner wormhole is locally isometric to the 
Einstein-Rosen manifold with two tubes (which is its `double cover'), 
but their topologies obviously differ. This means that for the observer 
outside the apparent horizons these two data sets are indistinguishable. 
This is not quite true for the Einstein-Rosen and  Schwarzschild data, 
which are not locally isometric. Even without exploring the region 
inside the horizons (which anyway is rendered impossible by existing 
results on topological censorship~\cite{Friedman}) 
they slightly differ in their interaction energy and other geometric
quantities, like e.g. the tidal deformation of the apparent horizons.
 
The two parameters $c$ and $\mu_0$ now label the two-hole configurations 
of equal mass. (In the Schwarzschild case the two independent parameters 
were $a\equiv a_1=a_2$ and $r_{12}\equiv\vert\vec x_1-\vec x_2\vert$.) 
But unlike the Schwarzschild case, we can now give not only closed 
analytic expressions for the total mass $M$ and individual mass $m$ 
in terms of the two parameters, but also for the geodesic distance of 
the apparent horizons $\ell$.  ($\ell$ is used as definition for 
`instantaneous distance of the two black holes'; for the Misner 
wormhole, where the two apparent horizons are identified, this 
corresponds to the length of the shortest geodesic winding once 
around the wormhole.) These read
\begin{equation}
M=4c\sum_{n=1}^{\infty}\frac{1}{\sinh n\mu_0},
\quad
m=2c\sum_{n=1}^{\infty}\frac{n}{\sinh n\mu_0},
\quad
\ell=2c(1+2m\mu_0).
\label{eq:2BH-ER-parameters}
\end{equation}
You might rightly wonder what `individual mass' should be if there 
is no internal end associated to each black hole where the ADM-formula 
(\ref{eq:ADM-mass}) can be applied. The answer is that there are 
alternative definitions of `quasi-local-mass' which can be applied 
even without asymptotic ends. The one we used for the expression 
of $m$ above is due to Lindquist \cite{Lindquist 1963} and is easy to 
compute in connection with the method of images. But it is lacking a 
deeper mathematical foundation. An alternative which is mathematically 
better founded is due to Penrose \cite{Penrose}, which however is much 
harder to calculate and only applies to a limited set of situations
(it agrees with the ADM mass whenever both definitions apply).
Amongst them are, however, all time-symmetric conformally flat data,
and for the data above the Penrose mass has fortunately been calculated 
in \cite{Tod}. The expression for $m$ is rather complicated and differs 
from that given above, though the difference is only of sixth order in 
an expansion in (mass/distance) \cite{Giulini 1990}. 
 
In summary, we see that the problem of setting up initial data 
for two black holes of given individual mass and given separation 
has no unique answer. Metrically as well as topologically different 
data sets can be found which have the same right to be called a 
realization of such a configuration. For holes without associated 
asymptotically flat ends no unambiguous definition for quasi-local 
mass exists.

\subsection{Non time-symmetric data}
According to a prescription found by Bowen and York~\cite{York 1979},
we can add linear and angular momentum within the setting of 
maximal data. We can still use \emph{conformally flat} data, i.e.
set $\hat{h}_{ij}=\delta_{ij}$, on multiply punctured $\reals^3$. 
Then the following two expressions add linear momentum $P^i$ and 
spin angular momentum $S^i$ to the puncture $\vec x=\vec 0$:
\begin{eqnarray}
\hat{K}_P^{ij}&=&\frac{3}{2r^2}
\left(P^i n^j+ P^j n^i-(\delta^{ij}-n^i n^j)
(\vec P\cdot \vec n)\right)\,,
\label{eq:K-momentum}\\
\hat{K}_S^{ij}&=&\frac{3}{r^3}((\vec S\times\vec n)^in^j+
                               (\vec S\times\vec n)^jn^i)\,.
\label{eq:K-angular-momentum}
\end{eqnarray}
It is straightforward to check that these expressions satisfy 
(\ref{eq:hat-constraints}) (note that $\hat{D}_i=\partial_i$). 
One can also check that these data will indeed give the proposed 
momenta and angular momenta at infinity (i.e. at the end 
$r\rightarrow\infty$). 
For this one may just use the `hatted' quantities in 
(\ref{eq:ADM-momentum}) and (\ref{eq:ADM-angular-momentum}), 
since the rescaling (\ref{eq:K-solution}) does not influence the 
leading order parts in the $1/r$ expansion of $K$, which alone 
contribute to these integrals. Linearity of all these 
equations in $K$ allows us to just add $K_P$ and $K_S$ and get 
initial data for one black hole with given momentum $\vec P$ and 
(spin) angular momentum $\vec S$. Moreover, we can add any finite 
number of expressions of the kind (\ref{eq:K-momentum} and 
\ref{eq:K-angular-momentum}) with parameters ${\vec P}_i,{\vec S}_i$ 
based at the puncture $\vec x_i$, where $i=1,\cdots, n$. This then 
leads to a data set 
whose total linear and angular momentum is given by the sum 
$\sum_i{\vec P}_i$ and $\sum_i{\vec S}_i$ respectively. But one 
may not immediately conclude that the ${\vec P}_i$ and ${\vec S}_i$
are linear and angular momenta of the individual black holes. 
Rather, the latter must be calculated for the internal ends of the 
manifold and for this one needs to know $\Phi$. 
The task then remains to solve  (\ref{eq:Lichnerowicz}) for the 
conformal factor, with blow-ups being allowed at the given punctures.

One interesting idea to facilitate solving (\ref{eq:Lichnerowicz}) is 
to first split off the singular part of $\Phi$, which blows up at the 
punctures $\{\vec x_1,\cdots,\vec x_n\}$ like 
$1/\vert\vec x-\vec x_i\vert$, from the regular 
reminder~\cite{Brandt} (compare also \cite{Beig}). One writes 
\begin{equation}
\Phi=\frac{1}{\alpha}+U,
\quad\mbox{with}\quad
\frac{1}{\alpha}:=\sum_{i=n}^n\frac{a_i}{\vert\vec x-\vec x_i\vert}\,,
\label{eq:split-Lichnerowicz}
\end{equation}
where the $a_i>0$ may be freely prescribed. Inserting this into 
(\ref{eq:Lichnerowicz}) gives 
\begin{equation}
\Delta U +\beta (1+\alpha U)^{-7}=0,
\quad\mbox{with}\quad 
\beta=\eights\alpha^7 K^{ij}K_{ij}\,.
\label{eq:BB-equation}
\end{equation}
The point is now the following: for $\vec x\rightarrow\vec x_i$
the function $\alpha$ tends to zero as $\vert\vec x-\vec x_i\vert$;
hence $\beta$, too, tends to zero as $\vert\vec x-\vec x_i\vert$.
This means that (\ref{eq:BB-equation}) has continuous coefficients
\emph{everywhere} in $\reals^3$ since the 
$1/\vert\vec x-\vec x_i\vert^6$--singularity at $\vec x_i$ of the 
$K$-squared term is cancelled by multiplication with $\alpha^7$. 
(Note that this relies on using the $K$'s from 
(\ref{eq:K-momentum}, \ref{eq:K-angular-momentum}), which possess 
no $1/r^n$ terms with $n>3$.) This means that equation 
(\ref{eq:BB-equation}) for $U$ can be solved on all of $\reals^3$,
without the need to excise the points $\{\vec x_1,\cdots,\vec x_n\}$ 
and therefore without the need to specify `inner' boundary conditions
for $U$; only the `outer' boundary-condition 
$U(r\rightarrow\infty)\rightarrow 1$ remains. This simplification 
seems particularly useful in numerical implementations 
(compare \cite{Brandt}).   

The total mass of our configuration is $M=\sum_i2a_i$. The individual 
masses are determined just as in \ref{subsubsec:two-BH}, by introducing 
the inverted radial coordinate $\bar{r}_i=a_i^2/r_i$ and reading off the 
coefficient of the $1/2\bar{r}_i$ term in the 
$\bar{r}_i\rightarrow\infty$ expansion. One easily gets 
\begin{equation}
m_i=2a_i(U(\vec x_i)+\chi_i)
\label{eq:boosted-masses}
\end{equation}
with $\chi_i$ as in (\ref{eq:mass-init}). 

The linear and angular momenta at, say, the $k$th end can also 
be calculated by using inverted coordinates, given by 
$\vec{\bar{x}}=(\vec x-{\vec x}_k)a^2_k/r^2_k$.
Expressed in these coordinates, the `hatted' (unphysical) extrinsic 
curvature tensor is given by $J^i_kJ^j_l\hat{K}_{ij}$ where
$J^i_k:=\frac{\partial x^i}{\partial{\bar{x}^k}}
=(a^2_k/\bar{r}^2)R^i_k$ with $R^i_k=\delta^i_k-2n^in_k$,
which is an orthogonal matrix. The `physical' extrinsic curvature 
is then obtained by multiplication with $\Phi^{-2}$ (compare 
\ref{eq:K-solution}). Now, 
$\Phi(\vec{\bar{x}})=\frac{\bar{r}}{a_k}(1+\frac{m_k}{2\bar{r}}+
O((1/\bar{r})^2))$
so that 
\begin{equation}
\bar{K}_{ij}=\left\{\bigl(
\frac{a_k}{\bar{r}}\bigr)^6+\mbox{terms}\propto
\bigl(\frac{a_k}{\bar{r}}\bigr)^p
\right\}
(R^k_iR^l_j\hat{K}_{kl}),\quad\mbox{where}\quad p\geq 7\,.
\label{eq:K-end-falloff}
\end{equation}
Inserting the expression (\ref{eq:K-momentum}) for $\hat{K}_P$
results in a $1/{\bar{r}}^4$ falloff so that the individual 
linear momenta are all zero \emph{as measured from the internal
ends}. One may say that the asymptotically flat internal 
ends represent the local rest frames of the black holes. Note that 
these rest frames are inertial since each black hole is freely falling. 
Inserting $K_S$ from (\ref{eq:K-angular-momentum}) gives a $1/{\bar{r}}^3$ 
falloff and an angular momentum which is just $-{\vec S}_k$ for the $k$th 
end. (Here one uses that $R^i_k$ is orientation-reversing orthogonal,
hence changing the sign of $\varepsilon_{ijk}$, and that 
$R^i_k(\vec S\times\vec n)^k=-(\vec S\times\vec n)^i$.)

\section{Problems and recent developments}
In this final section we draw attention to some of the current 
problems and developments, without claiming completeness.
\begin{enumerate}
\item[I.]
Given black hole data for $n$ holes of fixed masses and mutual 
separations (whatever definitions one uses here). One would like 
to minimize these data on the amount of outgoing radiation energy.
Any excess over the minimal amount can be said to be `already 
contained' initially. But so far no local (in time) criterion is 
known which quantifies the amount of gravitational radiation in an 
initial data set. First hints at the possibility that some 
(Newman-Penrose) conserved quantities could be useful here were 
discussed in \cite{Dain2}. 
\item[II.]
Restricting to spatially conformally flat metrics seems to be 
too narrow. It has been shown that there are no conformally flat
spatial slices in Kerr spacetime which are axisymmetric 
and reduce to slices of constant Schwarzschild time in the limit 
of vanishing angular momentum \cite{Garat 2000}. Accordingly,
Bowen-York data, even for a single black hole, contain 
excess gravitational radiation due to the relaxation of the individual 
holes to Kerr form \cite{Gleiser 1998}. See also \cite{Pullin GR15}
for an informal discussion of this and related problems. An alternative 
to the Bowen-York data which describe two spinning black holes and 
which reduce to Kerr data if the mass of one hole goes to zero 
have been discussed in~\cite{Dain1}.
\item[III.]
Even for the most simple 2-hole data (Schwarzschild or Einstein-Rosen) 
it is not known whether the evolving spacetime will have a suitably 
smooth asymptotic structure at future-lightlike infinity 
(i.e. `scri-plus'). As a consequence, we still do not know whether we 
can give a rigorous mathematical meaning to the notion of `energy loss 
by gravitational radiation' in this case of the simplest head-on 
collision of two black holes! The difficult analytical problems 
involved are studied in the framework of the so-called `conformal 
field equations'. See \cite{Friedrich GR15} (in particular section~4) 
for a summary and references.
\item[IV.]
We usually like to ask `Newtonian' questions, like: given two
black holes of individual masses $m_{1,2}$ and mutual separation 
$\ell$, what is their binding energy? For such a question to 
make sense we need good concepts of \emph{quasi-local mass}
and \emph{distance}. But these are ambiguous concepts in GR. 
Different definitions of `quasi-local mass' and `distance' amount 
to differences in the calculated binding energies which can be a few 
$10^{-3}$ times total energy at closest encounter \cite{Giulini 1990}. 
This is of the same order of magnitude as the total energy lost into 
gravitational radiation found in \cite{Baker 2000} for head-on 
(i.e. zero angular momentum) collision of two black holes modelled 
with Misner data. 
\end{enumerate}

\section{
Appendix: Eq. (\ref{eq:mod-poisson-eq}) satisfies the energy principle
\label{sec:appendix1}}
By the `energy principle' we understand the property, that \emph{all}
energy of the self-gravitating system serves as source for the 
gravitational field. In this appendix we wish to prove that 
(\ref{eq:mod-poisson-eq}) indeed satisfies this principle. 
For the uniqueness argument see \cite{Giulini 1996}.

Given a matter distribution $\rho$ immersed in its own gravitational 
potential $\phi$. Suppose we redistribute the matter within a 
bounded region of space by actively dragging it along the 
flow lines of a vector field $\vec\xi$ which vanishes outside some 
bounded region. The rate of change, $\delta\rho$, of the matter distribution 
is then determined through $\delta\rho\,dV=-L_{\vec\xi}\,(\rho\,dV)
=-\vec\nabla\cdot(\vec\xi\rho)\,dV$, where $L_{\vec\xi}$ is the 
Lie derivative with respect to $\vec\xi$ and $dV$ is the standard spatial 
volume element. Note that the latter also needs to be differentiated 
along $\vec\xi$, resulting in $L_{\vec\xi}\,dV=\vec\nabla\cdot\vec\xi$.
Hence we have $\delta\rho=-\vec\nabla\cdot(\vec\xi\rho)$.
The rate of work done to the system during this process is 
\begin{equation}
\delta A=\int_{\reals^3}dV\,\rho\vec\xi\cdot\vec\nabla\phi
        =-\int_{\reals^3}dV\,\phi\,\vec\nabla\cdot(\vec\xi\rho)
        =\int_{\reals^3}dV\,\phi\,\delta\rho\,,
\label{eq:app1}
\end{equation}
where the integration by parts does not lead to surface terms due to 
$\vec\xi$ vanishing outside a bounded region. Equation (\ref{eq:app1})
is still completely general, that is, independent of the field equation 
for $\phi$.  The field equation comes in when we assume that the 
process of redistribution is carried out adiabatically, which means that 
at each stage during the process $\phi$ satisfies its field equation
with the instantaneous matter distribution.
Our claim will be proven if under the hypothesis that $\phi$ satisfies 
(\ref{eq:mod-poisson-eq}) we can show that $\delta A=c^2\delta M_G$,
where $M_G$ is defined in (\ref{eq:grav-mass}) and represents the 
total gravitating energy according to the field equation.
Setting $\sqrt{\phi/c^2}=\psi$ and using the more convenient equation 
(\ref{eq:linearization}), we have 
\begin{eqnarray}
\delta A 
&=& \frac{c^4}{2\pi G}\int_{\reals^3} dV\,
    \psi^2\delta\left[\frac{\Delta\psi}{\psi}\right]
  = \frac{c^4}{2\pi G}\int_{\reals^3} dV\,
    \left[\psi\Delta(\delta\psi)-(\Delta\psi)\delta\psi\right]
    \nonumber\\
&=& \frac{c^4}{2\pi G}\,\lim_{r\rightarrow\infty}\int_{S^2(r)} d\sigma\,
    \vec n\cdot\left[\psi\vec\nabla(\delta\psi)-
                    (\vec\nabla\psi)\delta\psi\right]\,.
\label{eq:app2}
\end{eqnarray}
Now, the fall-off condition for $r\rightarrow\infty$ imply that 
$\vec\nabla\psi$ falls off as fast as $1/r^2$ and $\delta\psi$ as 
$1/r$. Hence the second term in the last line of (\ref{eq:app2}) 
does not contribute so that we may reverse its sign. This leads to 
\begin{eqnarray}
\delta A&=&\frac{c^4}{2\pi G}\,\delta\,\lim_{r\rightarrow\infty}
         \int_{S^2(r)}d\sigma\,(\vec n\cdot\vec\nabla\psi)\psi
        =\frac{c^4}{4\pi G}\,\delta\,\lim_{r\rightarrow\infty} 
         \int_{S^2(r)}d\sigma\,\vec n\cdot\vec\nabla\phi\nonumber\\
        &=& c^2 \delta M_G
\label{eq:app3}
\end{eqnarray}
which proves the claim.


\end{document}